\documentclass[14pt,a4paper]{extarticle}

\usepackage[utf8]{inputenc}
\usepackage[english]{babel}
\usepackage[T2A]{fontenc}

\usepackage{amsmath}
\usepackage{amsfonts}
\usepackage{amssymb}
\usepackage{graphicx}
\usepackage{hyperref}
\usepackage{color}
\usepackage{fullpage}
\usepackage{indentfirst}
\usepackage{gensymb}

\definecolor{darkblue}{RGB}{23,26,73}
\definecolor{darkgreen}{RGB}{27,37,14}
\definecolor{darkred}{RGB}{37,14,14}

\hypersetup{%
  colorlinks=true,%
  linkcolor=darkred,% Redefine this color to change links color
  citecolor=darkgreen,% Redefine this color to change cites color
  urlcolor=darkblue,% Redefine this color to change URIs color
  pdfauthor={N. V. Volkov, A. A. Lagutin, A. I. Reviakin, and R. T. Bizhanov},%
  pdftitle={Lateral Distribution Function of Extensive Air Showers Cherenkov Light and Stable Laws: Fast Modelling Method for the CORSIKA Code},%
  pdfsubject={Manuscript of an article for the journal BRAS},%
  pdfkeywords={Cosmic rays, extensive air showers, electromagnetic component, Cherenkov light, lateral distribution function, one-dimensional fractional stable distribution, CORSIKA code},%
  breaklinks=true,%
  plainpages=false%
}

\begin{document}

\begin{center}
{\large\bf Lateral Distribution Function of Extensive Air Showers Cherenkov Light and Stable Laws: Fast Modelling Method for the CORSIKA Code}

\vspace*{1cm}

N. V. Volkov\footnote{\url{https://orcid.org/0000-0002-3172-0655}}, A. A. Lagutin\footnote{\url{https://orcid.org/0000-0002-1814-8041}}, A. I. Reviakin, and R. T. Bizhanov

\vspace*{5mm}

Altai State University, Barnaul, 656049 Russia

\vspace*{5mm}

e-mail: \verb|volkov@theory.asu.ru|

\end{center}

\begin{abstract}

The paper proposes a new approach for approximating the lateral distribution functions (LDF) of Cherenkov light emitted by the electromagnetic component of extensive air showers (EAS) in the Earth’s atmosphere. The information basis of the study is a series of simulations with the CORSIKA code. To approximate the LDF atmospheric Cherenkov light the probability density functions of one-dimensional fractional stable distributions were used. The results obtained in the work allow us to propose a fast modeling method for the CORSIKA code using a procedure similar to the Nishimura-Kamata-Greisen (NKG) for calculating the LDF of the EAS electromagnetic component.

\vspace*{5mm}


\end{abstract}

\section*{Introduction}

At the present time in the high-energy astrophysics the main method for studying cosmic rays (CR) in high ($E>10^6$~GeV) and ultra-high ($E>10^{10}$~GeV) energies regions is the EAS method. The most important results of the last two decades on the spectrum and mass composition of CR were obtained in the largest experiments of the Pierre Auger Observatory [1], IceCube$\&$IceTop [2], Telescope Array [3], HAWC observatory [4], KASCADE-Grande [5], LHAASO [6], TAIGA gamma observatory [7] etc. Ground-based EAS arrays contain, as key segments, a system of scintillation detectors and/or optical telescopes that record Cherenkov light emitted by relativistic particles (primarily electrons and positrons) moving in the Earth’s atmosphere. Analysis of the Cherenkov light lateral distribution (radial density of Cherenkov photons as a function of distance to shower core) is a principle stage in the reconstruction of both CR energy spectrum and mass composition. One of the main methods of the LDF analysis is the selection of fitting functions.

The aim of this work is to develop a new approach for approximating the LDF atmospheric Cherenkov light emitted by the electromagnetic component of the EAS in the Earth’s atmosphere, using the results of the theory of stable distributions.

\section{LDF approximation}

Any function used to approximate the Cherenkov light lateral distribution must describe the main features of this distribution, namely power-law behavior of distribution before and after a certain point, so-called inflection point or break one.

The main function for approximating the LDF Cherenkov light and for reconstructing the EAS parameters in the TAIGA experiment [7] is the piecewise approximating function <<Tunka>> fit. The expression for this function and the approximation parameters are given for example in [8]. In the paper [9] a simpler five parameters approximation function <<knee-like>> fit is proposed. The <<knee-like>> fit was used in [10] to approximate the LDF obtained in the SPHERE-2 experiment. However, the <<knee-like>> fit showed less efficiency than the alternative approach using the sum of two polynomials with weight functions with exponential cutoff, suggested by the authors of [10].

In this paper it is proposed to use probability density functions of stable distributions as model functions to approximate the LDF atmospheric Cherenkov light from the EAS.

Stable distributions (SD) appear when summing independent and identically distributed random variables. If $\xi_1, \xi_2, \dots, \xi_n$ are independent random variables that have the same distribution, then the new value
$$\eta = \dfrac{1}{b}\sum_{i=1}^n \xi_i$$
at $n\rightarrow\infty$ will also be random with a distribution function that is a stable law [11,12]. Parameters $b$ and $a$ are the normalizing and centering constants, respectively. In the standard form ($b=1,a=0$), the SD are characterized by two main parameters $\alpha$ and $\beta$. The characteristic index $0<\alpha\leq 2$ determines the type of distribution, and the index $0<\beta\leq 1$ determines its asymmetry. The key feature of the SD is the absence of both the second moment at $1<\alpha<2$ and the first moment in one-sided distributions at any values of $\beta<1$. Special cases of the SD are the Gaussian ($\alpha=2, \beta=1$), Cauchy ($\alpha=1, \beta=1$), Levy (one-sided distribution at $\beta=1/2$), etc.

To approximate the Cherenkov light LDF, the paper uses the probability densities of the one-dimensional fractionally stable distribution $\Psi_1^{(\alpha,\beta)}(r)$ for different values of the characteristic indices $\alpha$ and $\beta$ [13]. The choice of this family of distributions is due to the presence of characteristic features: a break point and power-law asymptotes before and after this point [13]. We note that these important features are present in the LDF of the Cherenkov photons. In addition, we note that the authors have extensive experience working with the SD that arise in solutions of the CR diffusion equations in the Galaxy within the framework of the non-classical diffusion model created at the Altai State University (see, for example, [14-17]). Below in the paper the term <<SD-fit>> will be used for the proposed approximation method.

The SD-fit problem is written as
\begin{equation}
F=\underset{\{\alpha,\beta,F_0,R_{\text{Ch}}\}}{\min}\sum_{i=1}^n\left[Q(r_i) - F_0\Psi_1^{(\alpha,\beta)}\left(\dfrac{r_i}{R_{\text{Ch}}}\right)\right]^2
\end{equation}
Here $Q(r_i)$ is the density of Cherenkov photons at the distance $r_i$ from the shower core, $F_0$ is the scaling coefficient, $R_{\text{Ch}}$ is the break point of the LDF. The optimal values of parameters of the $F$ functional (1) were found using the Levenberg-Marquardt algorithm.

\section{Calculations}

To obtain the approximation parameters of the LDF atmospheric Cherenkov light, a series of simulations were carried out using the CORSIKA code version 7.7500 (April 2023) [18]. Calculations were made of the LDF from the EAS initiated by primary gamma quanta, protons and nuclei of helium, carbon and iron with energies of $10^4-10^9$~GeV, arrival angles of $0\degree$, $15\degree$ and $30\degree$ and depths of the first interaction in the atmosphere of $0$, $55$, $260$ and $550$~g/cm$^2$. As a result of calculations, a database was created containing in the form of multidimensional arrays the parameters $\alpha$, $\beta$, $F_0$ and $R_{\text{Ch}}$ for six values of the primary particle energy, for four values of the Cherenkov photon emission depth in the atmosphere and for three values of the primary particle arrival angles. The parameters of the SD for intermediate values can be found by interpolation of table values.

Figure 1a shows an approximations of the CORSIKA LDF of the Cherenkov light from vertical EAS initiated by primary proton and iron nucleus with energies $E=10^6$~GeV using SD. Figure 1b shows an example of the SD-fit of experimental data of the TAIGA-HiSCORE array. Table 1 shows the SD-fit parameters established as a result of the approximation of the CORSIKA proton showers.

\begin{figure}[htb]
\includegraphics[width=.5\textwidth]{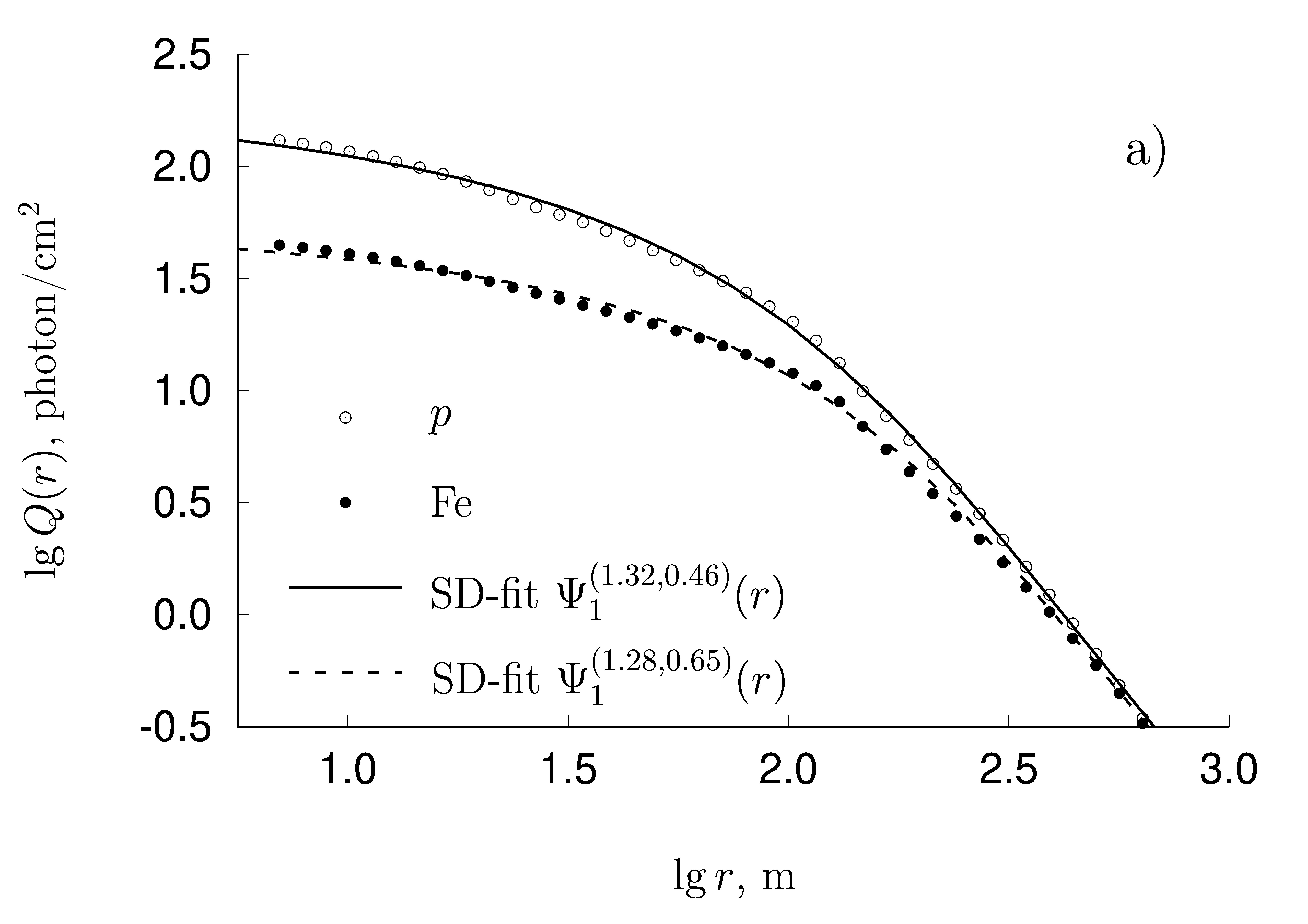}
\includegraphics[width=.5\textwidth]{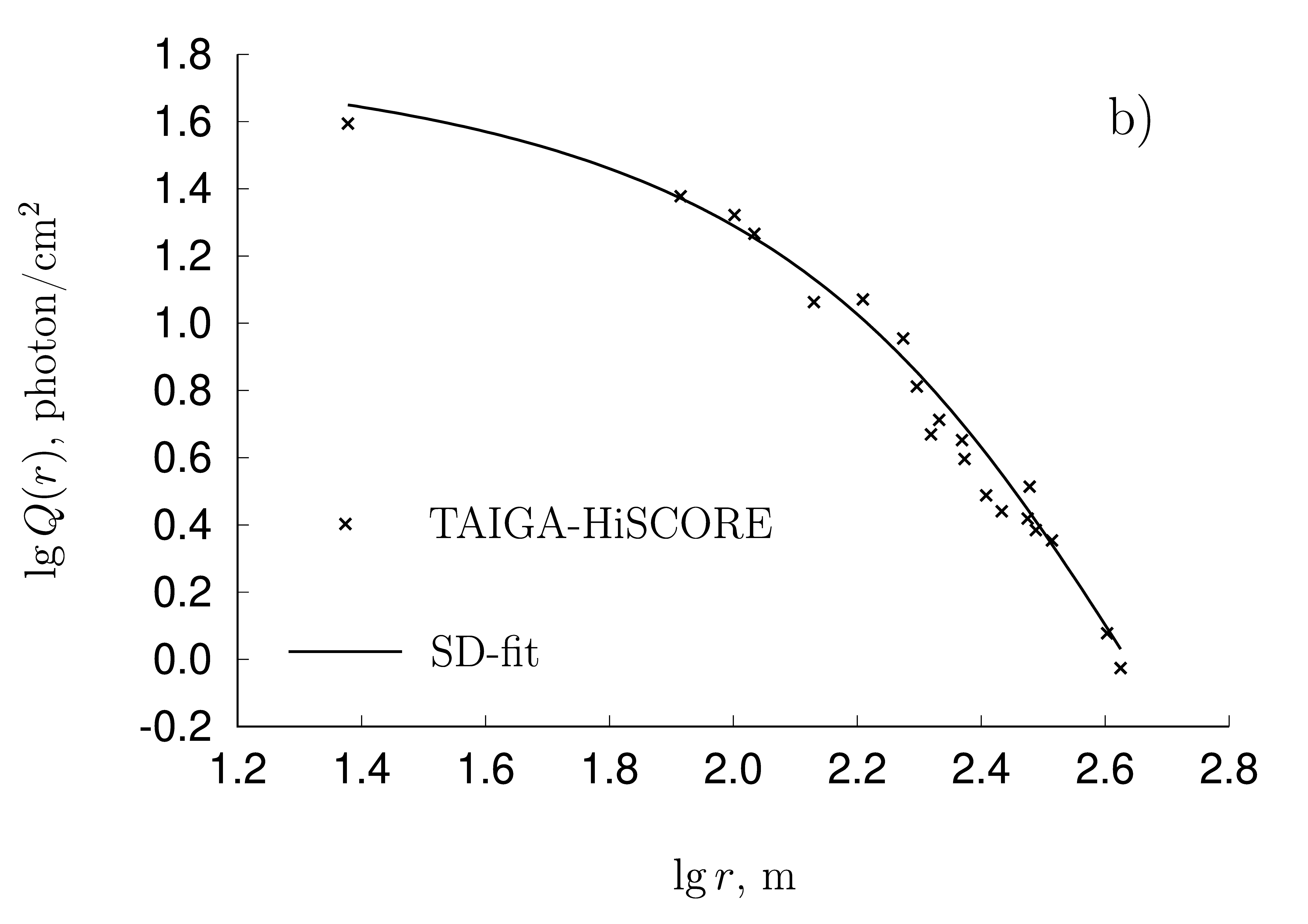}
\caption{SD-fit of the Cherenkov light LDF: a) CORSIKA vertical EAS initiated by primary protons (open circles) and iron nucleus (filled circles) with energies $E=10^6$~GeV, approximation parameters have the following values ($\alpha=1.32\pm 0.012, \beta=0.46\pm 0.007, \lg F_0=2.46\pm 0.12, R_{\text{Ch}}=52.6\pm 0.79$ for protons and $\alpha=1.28\pm 0.014, \beta=0.65\pm 0.008, \lg F_0=1.99\pm 0.13, R_{\text{Ch}}=78.1\pm 1.23$ for iron nucleus); b) SD-fit of the TAIGA-HiSCORE experimental data given in the paper [9] (crosses) with parameters $\alpha=1.52\pm 0.027, \beta=0.63\pm 0.020, \lg F_0=2.15\pm 0.24, R_{\text{Ch}}=81.20\pm 2.67$.}
\end{figure}

\begin{table}[hb]
\caption{SD-fit parameters of the CORSIKA Cherenkov light LDF from protons vertical EAS for different energies}
\begin{center}
\begin{tabular}{*{5}{|c}|}
\hline
$E$, GeV & $\alpha$ & $\beta$ & $\lg F_0$ & $R_{\text{Ch}}$ \\
\hline
$10^4$ & $1.21$ & $0.75$ & $3.90$ & $81.1$ \\
$10^5$ & $1.30$ & $0.53$ & $5.23$ & $60.2$ \\
$10^6$ & $1.32$ & $0.46$ & $6.46$ & $52.6$ \\
$10^7$ & $1.33$ & $0.27$ & $7.69$ & $42.2$ \\
$10^8$ & $1.33$ & $0.19$ & $8.88$ & $35.3$ \\
$10^9$ & $1.33$ & $0.11$ & $10.02$ & $32.4$ \\
\hline
\end{tabular}
\end{center}
\end{table}

\section{Fast simulation mathod for CORSIKA code}

The proposed new approach for approximating the LDF and created database of the SD parameters stimulated the development of a fast modeling method for the CORSIKA code. The main idea is to abandon the direct Monte Carlo modeling of electron-photon cascades. It is proposed to use the existing NKG module of the CORSIKA code which allows for fast calculations of lateral distribution of the EAS electron component using the modified Nishimura-Kamata-Greisen function [19]. The authors made changes to the source code of the NKG module. This module was supplemented with a procedure for calculating the LDF of Cherenkov photons using one-dimensional fractional-stable distributions according to a prepared database of parameters of these distributions.

\section{Break of the Cherenkov light LDF and mean square radius of electrons in EAS}

In the late 1990s, within the framework of the model-independent scaling approach to describing the electron LDF of EAS, a very simple approximation formula was proposed at the Altai State University to describe the mean square radius of electrons [20]
\begin{equation}
R_e(t) = \dfrac{\rho_0}{\rho(t)}173.0\left[0.546+\dfrac{2}{\pi}\text{arctg}\left(\dfrac{t}{t_{\max}+100 \text{g/cm}^2}-1\right)\right],\quad \text{m}.
\end{equation}
In Eq. (2) $\rho(t)$ is air density at depth $t$, $\rho_0=1.225$~g/cm$^3$. 

In this paper, using the results of simulations with the CORSIKA code, we carried out a study of the correlation between the mean square radius of electrons $R_e$ and the break of the LDF of Cherenkov light $R_{\text{Ch}}$. Figure 2 shows the dependence $R_e$ from $R_{\text{Ch}}$. It has been shown that $R_e=(0.77\pm 0.03)R_{\text{Ch}} + (83.11\pm 2.34)$.

To summarize, we note that the proposed SD-fit, as well as the established correlation between $R_e$ and $R_{\text{Ch}}$, allows us to obtain estimates of $R_e(R_{\text{Ch}})$ based on the approximation of experimental LDF, and this in turn allows us to obtain experimental estimates of depth of maximum $t_{\max}$. This result opens up the possibility of solving the problem of restoration of the primary cosmic rays mass composition.

\begin{figure}[htb]
\begin{center}
\includegraphics[width=.9\textwidth]{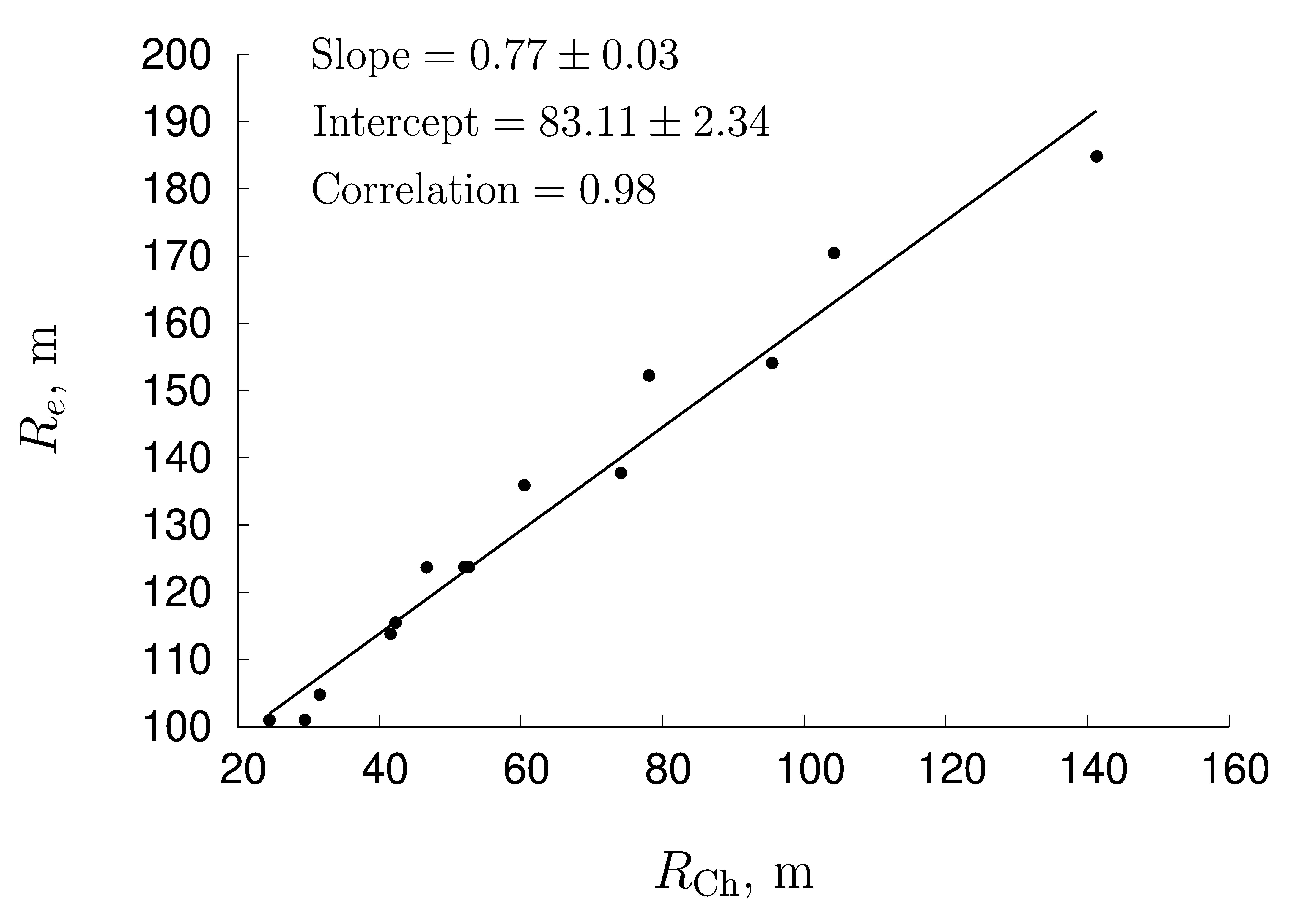}
\end{center}
\caption{Correlation between the mean square radius of electrons $R_e$ and the break of the LDF of Cherenkov light $R_{\text{Ch}}$.}
\end{figure}

\section*{Conclusion}

A new approach has been proposed for approximating the LDF of the atmospheric Cherenkov light emitted by the electromagnetic component of the EAS in the Earth’s atmosphere, using the results of the theory of stable distributions. SD-fit allows us to take into account the key features of the LDF when describing break point and power-law asymptotes before and after this point.

The results obtained in the framework of the new approach allow us to propose a fast modeling method for the CORSIKA code using a procedure similar to NKG for calculating the LDF of the EAS electromagnetic component.
Based on the simulation data from the CORSIKA code and the results of the scaling approach to describing the electron LDF of EAS, a correlation was established between the mean square radius of electrons $R_e$ and the break of the LDF of Cherenkov light $R_{\text{Ch}}$. It should be noted that this correlation allows us to solve one of the main problems of high-energy astrophysics: the restoration of the mass composition of primary cosmic rays.

\section*{Funding}

The work is supported by the Russian Science Foundation (grant no. 23-72-00057). 

\section*{References}
\begin{enumerate}
    \item J. Abraham, P. Abreu, M. Aglietta et al., Phys. Rev. Lett. 101, 061101 (2008). \url{https://doi.org/10.1103/PhysRevLett.101.061101}
    
    \item  M. G. Aartsen, M. Ackermann, J. Adams et al., Phys. Rev. D. 100, 082002 (2019). \url{https://doi.org/10.1103/PhysRevD.100.082002}
    
    \item  R. U. Abbasi, M. Abe, T. Abu-Zayyad et al., Astrophys. J. 865, 74 (2018). \url{https://doi.org/10.3847/1538-4357/aada05}
    
    \item  A. U. Abeysekara, A. Albert, R. Alfaro et al., NIMP A. 1052, 168253 (2023). \url{https://doi.org/10.1016/j.nima.2023.168253}
    
    \item  A. Haungs, J. C. Arteaga-Velazquez, M. Bertaina et al., Acta Phys. Polonica B Proc. Suppl. 15, 3-A2 (2022). \url{https://doi.org/10.5506/APhysPolBSupp.15.3-A2}
    
    \item  Z. Cao, F. Aharonian, Axikegu et al., Phys. Rev. Lett. 132, 131002 (2024). \url{https://doi.org/10.1103/PhysRevLett.132.131002}
    
    \item I. I. Astapov, P. A. Bezyazeekov, M. Blank et al., JETP. 134, 469 (2022). \url{https://doi.org/10.1134/S1063776122040136}
    
    \item V. V. Prosin, S. F. Berezhnev, N. M. Budnev et al., NIMP A. 94-101, 94 (2013). \url{http://doi.org/10.1016/j.nima.2013.09.018}
    
    \item A. S. M. Elshoukrofy, E. B. Postnikov, I. I. Astapov, Nucl. Phys. and Eng. 8, 311 (2017). In Russian. \url{http://doi.org/10.1134/S2079562917040091}
    
    \item V. S. Latypova, C. G. Azra, E. A. Bonvech et al., Proc of Sci. 27th ECRS. 423, 069 (2023).
     
    \item V. M. Zolotarev, One-dimensional stable distributions (Moscow, Nauka, 1983) (In Russian).
     
    \item V. M. Zolotarev, Stable laws and their applications (Moscow, Znanie, 1984). (In Russian). 
    
    \item V. V. Uchaikin and V. M. Zolotarev, Chance and Stability (VSP, Netherlands, Utrecht, 1999).
     
    \item A. A. Lagutin and N. V. Volkov, Bull. Russ. Acad. Sci. Phys. 85, 375 (2021). \url{http://doi.org/10.3103/S1062873821040213}
    
    \item A. A. Lagutin and N. V. Volkov, Phys. Atom. Nucl. 84, 975 (2021). \url{http://doi.org/10.1134/S1063778821130184}
    
    \item A. A. Lagutin, N. V. Volkov, and R. I. Raikin, Bull. Russ. Acad. Sci. Phys. 87, 878 (2023). \url{http://doi.org/10.3103/S1062873823702635}
    
    \item N. V. Volkov and A. A. Lagutin. Library of functions for calculating stable distributions. Statement about the state registration No. 2024662906. 
    
    \item D. Heck, J. Knapp, J. N. Capdevielle et al. CORSIKA: A Monte Carlo code to simulate extensive air showers. FZKA-6019. 1998.
    
    \item A. A. Lagutin, A. V. Pljasheshnikov, V. V. Uchaikin, Proc. 16th ICRC, 7, 18 (1979).
    
    \item R. I. Raikin, A. A. Lagutin, N. Inoue, and A. Misaki, Proc. 27th ICRC, 1, 294 (2001). 
    \end{enumerate}
\end{document}